\documentclass[12pt]{article}

\usepackage{amsmath,amsfonts}
\usepackage{amssymb}
\usepackage{graphicx}
\usepackage{epsfig,psfrag,cite}
\usepackage[usenames,dvips]{color}

\usepackage[margin=3.0cm]{geometry}

\makeatletter
\@addtoreset{equation}{section}
\makeatother

\newcommand{\be}{\begin{equation}}
\newcommand{\ee}{\end{equation}}
\newcommand{\bea}{\begin{eqnarray}}
\newcommand{\eea}{\end{eqnarray}}
\newcommand{\bml}{\begin{multline}}
\newcommand{\eml}{\end{multline}}

\begin{document}

\thispagestyle{empty}

\begin{center}
\hfill CPHT-RR044.0510

\begin{center}

\vspace{1.7cm}

{\LARGE\bf From Soft Walls to Infrared Branes}

\end{center}

\vspace{1.4cm}

{\bf Gero von Gersdorff$^{\;a}$}\\

\vspace{.1cm}

${}^a\!\!$ {\em {Centre de Physique Theorique, Ecole Polytechnique, CNRS\\
F-91128 Palaiseau, France}}

\end{center}

\vspace{0.8cm}

\centerline{\bf Abstract}
\vspace{2 mm}
\begin{quote}\small
Five dimensional warped spaces with soft walls are generalizations of the standard Randall-Sundrum compactifications, where instead of an infrared brane one has a curvature singularity (with vanishing warp factor) at finite proper distance in the bulk. 
We project the physics near the singularity onto a hypersurface located a small distance away from it in the bulk.
This results in a completely equivalent description of the soft wall in terms of an effective infrared brane, hiding any singular point. We perform explicitly this calculation for two classes of soft wall backgrounds used in the literature. 
The procedure has several advantages. It separates in a clean way the physics of the soft wall from the physics of the five dimensional bulk, facilitating a more direct comparison with standard two-brane warped compactifications. Moreover, consistent soft walls show a sort of universal behavior near the singularity which is reflected in the effective brane Lagrangian. Thirdly, for many purposes, a good approximation is obtained by assuming the bulk background away from the singularity to be the usual Randall-Sundrum metric, thus making the soft wall backgrounds better analytically tractable. 
We check the validity of this procedure by calculating the spectrum of bulk fields and comparing it to the exact result, finding very good agreement.
\end{quote}

\vfill

\section{Introduction}

The Randall Sundrum (RS) solution to the hierarchy problem \cite{Randall:1999ee} has by now become a serious competitor to supersymmetry for the physics of electroweak symmetry breaking (EWSB), predicting an interesting and peculiar TeV phenomenology.
In addition, with the AdS/CFT correspondence \cite{Maldacena:1997re,Witten:1998qj} we possess a dual interpretation of these theories in terms of four dimensional (4d) gauge theories. Although a good quantitative control over the duality only exists in the presence of supersymmetry, it can still be a very helpful tool in analyzing the plausibility of the 5d theory and provide us with some toy model to analyze the IR behavior of strongly coupled gauge theories. In the dual interpretation, the extra coordinate describes the renormalization group (RG) flow from the Planck to the TeV scale which is governed by a conformal fixed point.

In the classical setup, the 5d bulk is bounded by two 4d boundaries or branes, which are referred to as the UV and IR brane respectively. The natural cutoff for the theory becomes a function of the extra dimension, and in particular changes from the Planck scale (at the UV brane) to the TeV scale (at the IR brane). Throughout this paper, we will refer to the UV scale as $k$ and to IR scale as $\rho$ and have in mind that the former is of the order of the 5d Planck scale and equal to the AdS curvature, and the latter is of the order of a TeV.
The effective brane Lagrangians should thus be parametrized by a series of effective operators of higher and higher dimension, suppressed by their respective cutoff scale. 

The gauge couplings of the 5d theory are dimensionful quantities and hence should become strong at a scale somewhere above the TeV scale. By naive dimensional analysis this scale can be estimated to be of the order $\Lambda\sim \frac{12 \pi^3}{g_5^2}\frac{\rho}{k}\ll k$. One would like to trust the theory up to this scale in order to make meaningful predictions that could be verified at the LHC and other future colliders. However, this immediately causes a paradox: either one would need to specify an infinite number of operators on the IR brane or one cannot make any meaningful predictions for physics above but close to the TeV scale, i.e.~in the range $\rho<p<\Lambda$. In particular, KK masses cannot be predicted quantitatively, since an expansion of the IR brane Lagrangian in terms of $p^2$ is in general not convergent for $p>\rho$. The only way out is to resolve the IR brane.\footnote{
In fact, from the dual point of view the IR brane and its associated Lagrangian are somewhat mysterious objects. The IR brane provides IR boundary conditions on the RG flow, whereas one would expect that the latter is completely determined once we specify the boundary conditions in the UV (i.e., the UV brane Lagrangian).}
 Even though it is a thin (Planck-length) object, the warping makes it possible to probe its internal structure at TeV energies and hence it cannot be ignored.

In fact, the resolution of the IR brane can be realized quite naturally in terms of soft walls.
Note that in order to stabilize the hierarchy one needs to introduce some energy-momentum along the extra dimension, i.e., some scalar field that obtains a vacuum expectation value \cite{Goldberger:1999uk}. As a matter of fact, even in the absence of an IR brane, and under very general conditions (rather mild restrictions on the scalar potential in the bulk), the profile for the scalar field will diverge at finite proper distance $y_s$ in the bulk, driving the warp factor to zero. Once the metric vanishes, spacetime ends and we have dynamically generated a boundary. The point $y_s$ marks the location of a naked curvature singularity. 
In order to maintain the successful RS solution  to the hierarchy problem, it is sufficient to stabilize the position $y_s$ at about 30 Planck lengths $k^{-1}$ (i.e., about the same as the location of the IR brane in the two-brane scenario). This can be generically achieved without  fine tuning of the scalar potentials in the bulk and the UV brane~\cite{Cabrer:2009we}.
The warping can then be added on top of any stabilization mechanism (not affecting the location of $y_s$), and the IR scale $\rho$ will be suppressed by a warp factor $e^{-k y_s}$. As a matter of fact, for most of the bulk $0<y<y_s$, the metric will be almost RS like, while at a point $y_1<y_s$ the back-reaction of the scalar field will become important and drive it away from that solution.
We will refer to the region between $y_1$ and $y_s$ as the soft wall (SW). SW's have for instance been studied in the context of AdS/QCD \cite{AdS/QCD,Gursoy:2007cb,Batell1}, EWSB~\cite{Falkowski1,Batell2,Falkowski:2008yr,Cabrer2} and flavour physics\cite{Atkins:2010cc}.

From the dual point of view, the scalar field might be identified with the running gauge coupling of the strongly coupled $SU(N_c)$ gauge theory \cite{Gursoy:2007cb}, or, more generally, with some other relevant deformation of the CFT. For several e-folds $\sim ky_1$ of RG flow it stays very close to its fixed point, and then starts to diverge from it causing the theory to confine and generate a mass gap and a discrete spectrum of excitations. The SW background thus gives the precise quantitative behaviour of the the gauge theory away from the fixed point, whose knowledge is necessary in order to make quantitative predictions for the IR behaviour of the theory.

\begin{figure}[t]
\begin{center}
\includegraphics{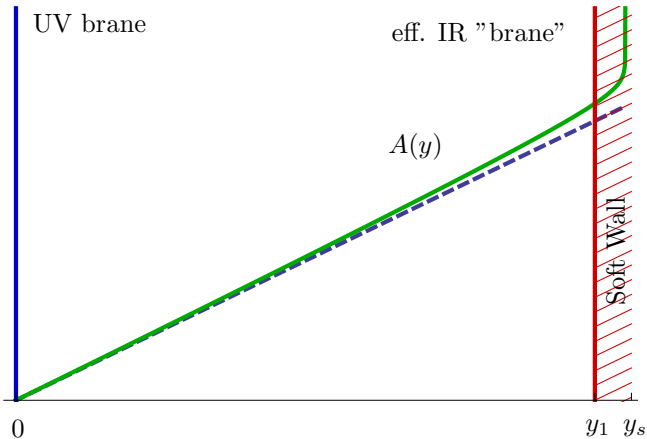}
\end{center}
\caption{Integrating over soft walls. The warp factor $A(y)$ starts to deviate from the RS solution $A(y)=k y$ at $y=y_1$ and diverges at $y=y_s$.
The region between $y_1$ and $y_s$ is integrated over and generates an effective IR brane at $y_1$. The plot is an actual representation of the SW1 model with $\nu=1.3$ and $k y_s=30$.}
\label{fig1}
\end{figure}

The purpose of this note is to study some prime examples of SW's and integrate over the region $y_1<y<y_s$, obtaining an effective IR Lagrangian at $y=y_1$, see Fig.~\ref{fig1}. The SW is thus replaced by an effective IR brane, allowing to incorporate the effects of the SW
in conventional two-brane setups, with the advantage that one now has some confidence of how the latter behave in the region $\rho<p<\Lambda$. 
Moreover, approximating the bulk background for $y<y_1$ with a pure RS metric defines a useful approximation scheme, given that in typical SW backgrounds  one rarely has full analytic control over the entire coordinate range.
The advertised procedure is essentially a matching of solutions to the equations of motion in the two regions $y<y_1$ and $y>y_1$, the fact that we parametrize the solution coming from $y>y_1$ as a Lagrangian at $y_1$ is strictly speaking not necessary, but kind of instructive in order to compare with hard wall models. On the other hand, this last step does not impose any further technical complications.

The rest of this paper is organized as follows. In Sec.~\ref{classes} we classify consistent SW backgrounds by their behavior near the singularity. In Sec.~\ref{Leff} we review the technique of holographic projection and apply it to a scalar field propagating near SW's. Finally in Sec.~\ref{spec} we apply our formalism to calculate the spectrum and compare it to exact numerical an analytical results. In Sec.~\ref{concl} we present our conclusions.

\section{Classes of Soft Walls}

\label{classes}

As mentioned in the introduction, the SW background is sourced by a scalar field $\phi$.
In order to simplify the scalar-gravity system, it is convenient to consider an integrated version of the scalar potential, the superpotential $W$ \cite{Brandhuber:1999hb,DeWolfe:1999cp}, defined by the differential equation~\footnote{We follow the convention of \cite{Cabrer:2009we}. Other works differ in the normalization of the superpotential and/or the field $\phi$. We work in units of the 5d Planck mass, i.e.~$M_5=1$.}
\be
V(\phi)=3W'(\phi)^2-12 W(\phi)^2\,.
\label{super}
\ee
Potentials of this kind appear in certain 5d gauged supergravities \cite{Freedman:1999gp}. Here, we will simply consider Eq.~(\ref{super}) as a definition for the auxiliary quantity $W$ that simplifies the Einstein equations. In fact, making the additional assumptions of 4d Poincar\'e invariance,\footnote{Our convention is $\eta_{\mu\nu}=(-+++)$. All metrics are understood to be in the 5d Einstein frame.} 
\be
ds^2=e^{-2A(y)}\eta_{\mu\nu}dx^\mu dx^\nu+dy^2\,,
\label{metricy}
\ee
the equations of motion (EOM) become
\bea
\phi'(y)&=&\frac{d}{d\phi}W(\phi)\,,\label{phi'}\\
A'(y)&=&W(\phi)\,.
\label{A'}
\eea
Introducing a boundary potential $\lambda(\phi)$ on the UV brane, the boundary value $\phi_0\equiv\phi(0)$ is determined by extremizing the 4d potential.
\be
V_{4d}(\phi)=\lambda(\phi)-6W(\phi)\,.
\ee
and demanding $V_{4d}(\phi_0)=0$. The latter condition ensures the vanishing of the 4d cosmological constant, needed to ensure consistency of our 4d flat ansatz for the metric.
Sometimes coordinates other than Eq.~(\ref{metricy}) are useful. In particular, we will also make use of the conformally flat coordinates 
\be
ds^2=e^{-2A(z)}\left(\eta_{\mu\nu}dx^\mu dx^\nu+dz^2\right)\,,
\label{metricz}
\ee
which are related to the "proper distance" coordinates defined in Eq.~(\ref{metricy}) by the differential equation $d y = e^{-A(z)} dz$.
Notice that one can write formal solutions to Eqns.~(\ref{phi'}), (\ref{A'}):
\be
A(\phi)=\int_{\phi_0}^\phi \frac{W(\phi')}{W'(\phi')}d\phi'
\label{formal1}
\ee
as well as
\be
y-y_0=\int_{\phi_0}^\phi \frac{1}{W'(\phi')}d\phi'\,,
\qquad
z-z_0=\int_{\phi_0}^\phi \frac{e^{-A(\phi')}}{W'(\phi')}d\phi'\,.
\label{formal2}
\ee

In fact, the class of realistic SW's is surprisingly small. If we insist that we have a mass gap and a meaningful curvature singularity \cite{Forste:2000ps,Gubser:2000nd,Cabrer:2009we}, the admissible scalar potentials have a restricted asymptotic behaviour at large field values. These asymptotic behaviours have been identified in Refs.~\cite{Gursoy:2007cb,Cabrer:2009we}.
In particular, a mass gap exists if $W(\phi)$ diverges as $e^\phi$ or faster, while the singularity can be made sense of only if it diverges more slowly than $e^{2\phi}$ \cite{Cabrer:2009we}.
We will consider two classes of SW's. Type-1 models (SW1) follow from a superpotential which asymptotically behaves as
\be
W(\phi)\sim e^{\nu\phi}\,,\qquad 1\leq \nu< 2\,,
\ee
at $\phi\to \infty$. Here we restrict to the case $\nu\geq 1$, which leads to a mass gap. The case $\nu=1$ is special. In this case the subleading behaviour actually becomes important, which leads us to type-2 Soft Walls (SW2),
\be
W(\phi)\sim e^{\phi}\phi^\beta\,,\qquad \beta> 0\,.
\ee
Note the requirement of the existence of a mass gap implies $\beta\geq 0$ with the special case $\beta=0$ already contained in the SW1 models.

Models with SW's typically have several energy scales which show up as different points along the extra dimension. Firstly, there is the UV cutoff of the theory, which we denote by $k$ and which we take to coincide with the AdS curvature scale. It corresponds to the location of the UV brane at $y_0=0$, or $z_0=k^{-1}$. Next
 there is the point along the extra coordinate where the back reaction of the scalar field becomes non-negligible. We will denote it with $y_1$ ($z_1$) and the corresponding energy scale with $\rho=z_1^{-1}$. In dual language, it is the RG scale at which the theory flows away from the conformal regime.  Finally, there is the point $y_s$ ($z_s$), at which the curvature singularity appears and spacetime ends. From Eq.~(\ref{formal2}) one can calculate the values of $y_s$ and $z_s$ by setting the upper integration limit to $\phi=\infty$. 
In both SW1 and SW2, $y_s$ (and hence $y_1<y_s$) are always finite. In SW1,  $z_1<\infty$,  and $z_s<\infty$ for $\nu>1$. In particular, $z_s^{-1}$ is the asymptotic spacing of the KK modes, or, equivalently the excitations of the gauge theory.
 In SW2, $z_1<\infty$ while $z_s<\infty$ for $\beta>\frac{1}{2}$. We will also define $\Delta_y =y_s-y_1$ and refer to it as the wall thickness. Note that the definition of $y_1$ is to a certain degree arbitrary. On the one hand we would like to have $y_1$ small enough such that the bulk $y<y_1$ is well approximated by RS. On the other hand, soft walls show a kind of universal behavior in the region $y$ very close to $y_s$, in the sense that the asymptotic properties of consistent soft walls are quite constrained.
This will become clearer when we turn to actual examples.

Before continuing, let us summarize the asypmtotic behaviour in the coodinates Eq.~(\ref{metricy}) and (\ref{metricz}).
SW1 backgrounds have the leading asymptotic form 
\be
A(y)=-\frac{1}{\nu^2}\log\left(1-\frac{y}{y_s}\right)\,,
\label{sw1pd}
\ee
near $y=y_s$ and
\be
A(z)=-\frac{1}{\nu^2-1}\log\left(1-\frac{z}{z_s}\right)\,,\qquad \nu>1\,,
\label{SW1zs}
\ee
near $z=z_s$.\footnote{For $\nu=1$ the behaviour is $A(z)=z+\dots$ as $z\to\infty$.}
The SW2 background also behaves as
\be
A(y)=-\log\left(1-\frac{y}{y_s}\right)\,,
\label{sw2pd}
\ee
but the asymptotic form in conformally flat coordinates now depends on the subleading behaviour of the superpotential, parametrized by the exponent $\beta$:
\be
A(z)\sim \left\{
\begin{array}{ll}
(\rho z)^{\frac{1}{1-2\beta}}&\beta<\frac{1}{2}\,, \\
e^{\rho z}&\beta=\frac{1}{2}\,,\\
(\rho[z_s-z])^\frac{1}{1-2\beta}&\beta>\frac{1}{2}\,.
\end{array}
\right.
\label{sw2cf}
\ee
At $\beta>\frac{1}{2}$, the location of the singularity becomes again finite, $z_s<\infty$. We will not consider this case here, as it is very similar to the SW1 models with $\nu>1$.

Although we have decribed here only dynamical SW's generated by the profile of a single field, the question of whether a given SW background is acceptable (has a mass gap and defines a meaningful singularity) at the end only depends on the profile $A(y)$. In particular, the criterion for a good singularity, Eq.(2.16) of Ref.~\cite{Cabrer:2009we}, generalizes to several fields
\be
\lim_{y\to y_s} e^{-4A(y)}W(\phi_i(y))\sim\lim_{y\to y_s} \left(e^{-4A(y)}\right)'=0
\label{condcc}
\ee
Although this does not result in a simple criterion on the (super)potential as in the case of a single field, it again restricts the asymptotic behaviour of the metric near the singularity, leading to the same classification of SW's.
Notice that the models of Refs.~\cite{Brandhuber:1999hb,Cabrer:2009we,Falkowski:2008yr} fall in the SW1 category whereas the ones in Refs.~\cite{AdS/QCD,Batell1,Batell2,Falkowski1,Atkins:2010cc} correspond to SW2 models. All the models studied in Refs.~\cite{AdS/QCD,Batell1,Batell2,Falkowski1,Atkins:2010cc} belong to the case $\beta<\frac{1}{2}$ in our description, and only have $y_s<\infty$ but  $z_s=\infty$.

Let us now proceed to integrate over the region $y_1<y<y_s$ and calculate the effective IR-Lagrangian that describes the SW.

\section{The SW Effective Lagrangian}
\label{Leff}

Let us start by considering a scalar field in the SW background with action
\be
S=\int_0^{y_s} d^5x  \,\mathcal L_{\rm bulk}\,,\qquad 
\mathcal L_{\rm bulk}=\frac{1}{2}\sqrt{-g}\left( g^{MN}\,\partial_M\psi\partial_N\psi+M^2\psi^2\right)\,.
\ee
We would like to integrate this between $y_1$ and $y_s$, and rewrite $S$ as
\be
S=\int_0^{y_1} d^5x\, \left[\mathcal L_{\rm bulk}+\mathcal L_{\rm SW}\,\delta(y-y_1)\right]\,.
\label{action2}
\ee
Note that the bulk integration is now restricted to $y<y_1$, where we might approximate the background by a pure RS metric. The physics beyond $y=y_1$ is contained in the effective SW Lagrangian $\mathcal L_{\rm SW}$.

To integrate out properly the physics beyond $y_1$, one must solve the equations of motion as a function of the field value at $y=y_1$. 
Technically, we holographically project the phsyics at $y>y_1$ onto a hypersurface at $y=y_1$:~\footnote{There is a change of sign because the orientation of the hypersurface at $y=y_1$ depends on whether it is a viewed as a boundary of the space $y>y_1$ or $y<y_1$.}
\be
\sqrt{-g(y_1)}\,\mathcal L_{SW}=-\int_{y_1}^{y_s} \sqrt{-g}
\,\mathcal L^{\rm on-shell}\,,\label{holo}
\ee
where $\mathcal L^{\rm on-shell}$ is regarded as a functional of $\psi(x,y_1)$.
We proceed by calculating the kernel \cite{Witten:1998qj}
\be
\left(
e^{2A(y)} \partial_\mu\partial^\mu
+\partial_y^2
-4A'(y)\partial_y
-M^2
\right)
 K(x,x';y)=0\,,
 \label{eom}
\ee
with the boundary condition $K(x,x',y_1)=\delta(x-x')$. We can then write 
\be
\psi(x,y)=\int d^4x' K(x,x';y)\psi(x',y_1)\,,
\ee
to find the solution to the EOM for $\psi$ as a function of $\psi(y_1)$. To fully specify the problem, we need to demand that the fields at $y=y_s$ behave regularly which gives us another boundary condition on $K$. The equation for $K$, written in momentum space for the 4d coordinates, gives two linearly independent solutions. Calling $K^{\rm reg}$ the solution that is regular at $y=y_s$, the boundary condition at $y_1$ implies that
\be
K(p,y)=\frac{K^{\rm reg}(p,y)}{K^{\rm reg}(p,y_1)}\,.
\ee
Plugging this back into the action we find
\be
\mathcal S_{\rm SW} = \frac{1}{2}\int \frac{d^4p}{(2\pi)^4}\sqrt{-g(y)}\, \mathcal F(p)\, \psi(-p,y)\psi(p,y)\, \delta(y-y_1)\,,
\ee
where we have defined the 
"form factor" 
\be
\mathcal F(p)=K'(p,y_1)\,.
\ee
Notice that $\mathcal F(p)$ is the inverse of the propagator $\langle \psi (-p,y_1)\psi(p,y_1)\rangle$ that would result if $\psi$ obeyed Neumann boundary conditions at $y_1$. For some backgrounds, other coordinates are more convenient. in particular, the equation of motion for $K$ in conformally flat coordinates reads
\be
\left(
\partial_\mu\partial^\mu
+\partial_z^2
-3 A'(z)\partial_z
-e^{-2A(z)} M^2
\right)
 K(x,x';z)=0\,.
 \label{eomcf}
\ee
In the following we will compute $\mathcal F$ in the two classes of SW's introduced previously. 

\subsection{Soft Walls of type 1}

We will take the full metric as
\be
e^{-A(y)}=e^{-k y}\left(1-\frac{y}{y_s}\right)^{\frac{1}{\nu^2}}\,,\qquad 1\leq\nu<2\,.
\label{SW1exact}
\ee
This background corresponds to an exact solution to the Einstein equations and was extensively studied in Refs.~\cite{Cabrer:2009we} and \cite{Cabrer2}. The first factor can be attributed to the negative bulk cosmological constant. The second factor is the (exact) back reaction of a scalar field with a certain potential. The point $y_s$ marks the location of the singularity and is taken to be $ky_s\sim\mathcal O({30})$.

The definition of the scale $z_1$ and $y_1$ is a bit arbitrary. We will take it to be
\be
z_1= k^{-1}(k y_s)^{\frac{1}{\nu^2}}e^{k y_s}\,,\qquad 
z_s=\Gamma(1-\tfrac{1}{\nu^2})z_1\,,
\label{z1zs}
\ee
where for reference we also give the explicit expression for $z_s$.
Notice that $z_1<z_s$. It has been found in Ref.~\cite{Cabrer:2009we} that the spectrum resulting from such a background can be characterized by a mass gap of the order $\sim z_1^{-1}$ and an asymptotic spacing $\Delta m_n=\pi/z_s$ for $n$ large. Only for $\nu=1$ the location of the singularity $z_s$ goes to infinity in conformal coordinates (due to the pole of the $\Gamma$ function), resulting in a continuum of states above the gap.
Notice that near the singularity the change of variables is governed by the relation
\be
\rho(z_s-z)=\frac{\nu^2}{\nu^2-1}\left(1-\frac{y}{y_s}\right)^{1-\frac{1}{\nu^2}}\,,
\ee
which is a good approximation for $k (y_s-y)\lesssim 1$. With the above definition for $z_1$ it follows that $k \Delta_y\equiv k(y_s-y_1)\lesssim 1$ with a mild $\nu$ dependence.

The solution for the kernel $K$ in the background Eq.~(\ref{SW1exact}) can easily be calculated. In the regime $|p|\gg  M/(k z_1)$
it is useful to switch to conformally flat coordinates and solve Eq.~(\ref{eomcf}) in the approximation Eq.~(\ref{SW1zs}) to obtain:\footnote{Note that the Bessel functions encountered here have nothing to do with the Bessel functions found in the bulk of the RS background.}
\be
K^{\rm reg}(p,z)=(z_s-z)^{-\alpha} J_\alpha(\sqrt{-p^2} (z_s-z))\,,\qquad \alpha=\frac{4-\nu^2}{2(\nu^2-1)}
\label{SW1plarge}\,,
\ee
resulting in
\be
\mathcal F(p)=e^{A(z_1)}\,\frac{\sqrt{-p^2}\, J_{\alpha+1}(\sqrt{-p^2}\, \Delta_z)}{J_\alpha( \sqrt{-p^2}\, \Delta_z)}\,,
\label{largep}
\ee
where $\Delta_z=z_s-z_1$, the wall thickness in conformal coordinates (this is different from the phsyical brane thickness that is given by $\Delta_y=y_s-y_1$).

On the other hand, the regime of small momentum, $p\ll M/(k z_1)$ is accessible in the $y$-coordinates
\be
K^{\rm reg}(0,y)=(y_s-y)^{-\alpha'} I_{\alpha'}(M (y_s-y))\,,\qquad \alpha'=\frac{4-\nu^2}{2\nu^2}\,.
\ee
The $p=0$ part of the form factor, i.e.~the effective IR brane mass is thus given by
\be
M_{\rm IR}\equiv\mathcal F(0)=-M\,\frac{ I_{\alpha'+1}(M\, \Delta_y)}{I_{\alpha'}( M\, \Delta_y)}\,.
\ee
It is reassuring that the so calculated brane mass term is a pure effect of the original bulk mass term: Setting $M=0$ we indeed obtain $M_{\rm IR}=0$. Furthermore, if $M\sim\mathcal O(k)$, then $M_{\rm IR}\sim\mathcal O(k)$. 
Still, such a large brane mass does not cut off all the fluctuations of the field on the IR brane: As soon as $-p^2$ becomes bigger than $M^2 e^{-2A(y_1)}$, there is a nontrivial Lagrangian given by Eq.~(\ref{largep}). This reflects the fact that despite 
the very thin SW, $\Delta_y \sim\mathcal O(k^{-1})$, already much lower energies than $k$ can resolve it due to the warping.

Finally we calculate the form factor for the interesting case of $\nu=1$:
\be
\mathcal F(p)=\frac{\frac{3}{2}-\beta(p)}{ \Delta_y}-\frac{M I_{1+\beta(p)}(M\, \Delta_y)}{I_{\beta(p)}(M\, \Delta_y)}\,,\qquad
\beta(p)=\sqrt{\frac{9}{4}+\frac{p^2}{\rho^2}}\,.
\ee
Notice that for $-p^2>\frac{9}{4}\rho^2$, the form factor develops an imaginary part. This signals the onset of the continuum. In fact, the situation is analogous in ordinary 4d quantum field theory, where the inverse propagator (the form factor) at $-p^2>4m^2$ develops an imaginary part due to the possibility of the creation of a two-particle state, resulting in a continuum of states. 

\subsection{Soft Walls of Type 2}

Next we consider SW's with a metric
\be
A(z)=\log(kz)+(\rho z)^\frac{1}{1-2\beta}\,,\qquad 0\leq \beta< \frac{1}{2}\,,
\label{type2}
\ee
where 
$\rho=z_1^{-1}$ is given by  Eq.~(\ref{z1zs}) with $\nu=1$. Although Eq.~(\ref{type2}) as it stands does not follow from a simple scalar-gravity system, it has been shown in Ref.~\cite{Cabrer:2009we} that a metric with identical asymptotic properties for small and large $z$ can be constructed dynamically, where indeed the scale $\rho$ is naturally warped down with respect to $k$. Here, we will only use the asymptotic properties so we will not worry about the details of the full metric. Notice that $z_s=\infty$ for the whole range of $\beta$, and $z_1$ indeed marks the point at which the second term (the backreaction due to the scalar field) becomes important and finally dominates over the first term (the warping induced by the cosmological constant).

For generic $\beta$ it is difficult to obtain exact solutions near the singularity. However, a WKB approximation can be employed. To this end we rescale the scalar field as $\psi=e^{\frac{3}{2}A}\tilde \psi$ and write the Schr\"odinger equation
\be
\tilde\psi''(z)-p^2\tilde\psi(z)- V(z)\tilde\psi(z)=0\,,\qquad V(z)=\frac{9}{4}A'(z)^2-\frac{3}{2}A''(z)+e^{-2A(z)}M^2\,,
\ee
and calculate the regular solution in the WKB approximation as
\be
\tilde \psi^{\rm reg}(z)
=\frac{2}{(-p^2-V_s(z))^{1/4}}\cos\left(\int_z^{ z_2(p)}\sqrt{-p^2-V(z')}-\tfrac{\pi}{4}\right)\,,
\ee
where $z_2(p)>z_1$ defines the turning point $V(z_2)+p^2=0$.
Near the singularity (i.e., $z\to\infty$) one can approximate
\be
V(z)\approx \frac{9\, \rho^2}{4\,(1-2\beta)^2} (\rho z)^\frac{4\beta}{1-2\beta}\,,
\ee
and obtain the form factor in the limit $-p^2\gg\rho^2$
\be
\mathcal F(p^2)=
e^{A(z_1)}\sqrt{-p^2}\tan \left(c_\beta \left[-\frac{p^2}{\rho^2}  \right]^\frac{1}{4\beta}
\right)\,,
\label{FF2}
\ee
with the constant $c_\beta$ defined as
\be
c_\beta=\frac{3\sqrt{\pi}}{4}\frac{\Gamma(\frac{1}{4\beta}-\frac{1}{2})}{\Gamma(\frac{1}{4\beta})}
\left(\frac{2(1-2\beta)}{3}\right)^{\frac{1}{2 \beta}}\,.
\ee

\section{Spectrum}
\label{spec}

\begin{figure}[t]
\begin{center}
\includegraphics[width=9cm]{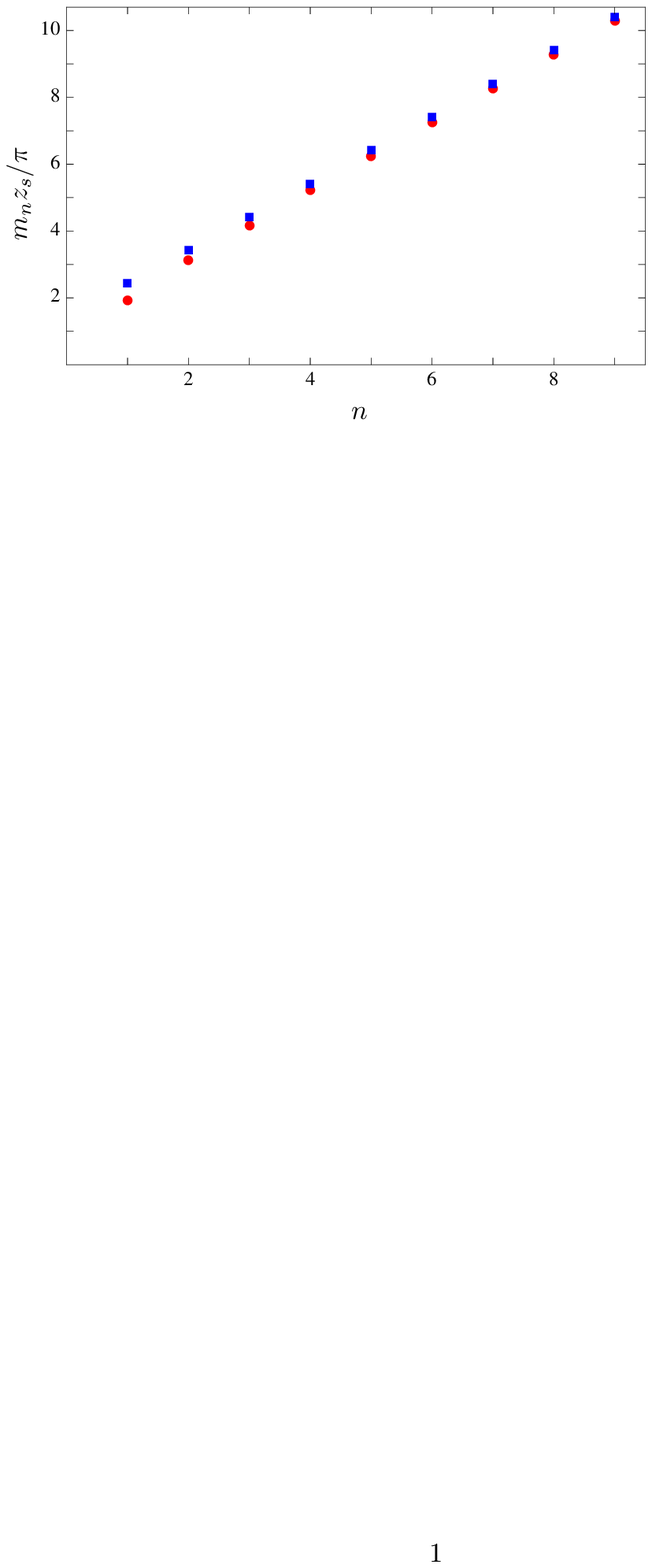}
\end{center}
\caption{Exact spectrum of the SW1 model (blue squares) and the approximation Eq.~(\ref{approx1}) (red circles) for $\nu=1.3$, $M=0$. Note the mas gap, i.e.~the lightest mode, is larger than the asymptotic spacing, a characteristic feature of the SW1 models. In this example the gap is about twice the spacing.}
\label{fig2}
\end{figure}

The spectrum of KK modes can be obtained from the effective IR Lagrangian as follows. Let us call $\psi_n$ a solution to the wave equation in the bulk that satisfies the boundary condition at $y=0$ ($z=z_0$). Then, variation of the action Eq.~(\ref{action2}) leads to the IR boundary condition at $y_1$ ($z_1$)
\be
\psi_n'(y_1)=\mathcal F(-m_n^2)\psi_n(y_1)\,,\qquad e^{A(z_1)}\psi_n'(z_1)=\mathcal F(-m_n^2)\psi_n(z_1)\,.
\label{BCeff}
\ee
In particular, for a scalar in a pure RS background with Dirichlet BCs at the UV brane one obtains the well known solution 
\be
\psi_n(z)=z^2\bigl[Y_{q}(m_n z_0)J_{q}(m_nz)-J_{q}(m_nz_0)Y_{q}(m_nz)\bigr]\,,\qquad q=\sqrt{4+M^2/k}\,,
\label{RSsol}
\ee
Then, using Eqns.~Eq.~(\ref{SW1plarge}), (\ref{BCeff}) and (\ref{RSsol}), for $m_n\ll k$ one has for the SW1 background
\be
\left(\frac{2-q}{z_1m_n}+\frac{J_{q-1}(m_nz_1)}{J_{q}(m_nz_1)}\right)=\frac{J_{\alpha+1}(m_n\Delta_z)}{J_{\alpha}(m_n\Delta_z)}\,.
\label{approx1}
\ee
We can compare these findings with the numerical result. The exact numerical eigenvalues~\cite{Cabrer:2009we} of a scalar in the background Eq.~(\ref{SW1exact}), with $M=0$ ($q=2$) and $\nu=1.3$, are plotted in Fig.~\ref{fig2}, together with those obtained with the form factor method, Eq.~(\ref{approx1}). The first modes are a little off due to the neglect of the subleading terms of the metric near the singularity. The deviation is  about 20\% for the lightest mode and about 1\% for the 10$^{\rm th}$ mode.
Furthermore, 
the asymptotic\footnote{Note that this is not a true asymptotic value for $n\to\infty$ but rather for the range $\rho\ll m_n\ll k$. For $m_n\gg k$ the asymptotic shift w.r.t.~$n$ can actually be computed as $\frac{\alpha}{2}-\frac{1}{4}$.} spectrum is 
\be
z_s m_n=\left(n+\frac{\alpha-q-1
}{2}\right)\pi\,.
\label{approx2}
\ee
Numerically, for large $n$ we obtain eigenvalues $m_nz_s/\pi=(10.40,\,50.37,\,100.36,\, 500.35)$ for $n=(10,\ 50,\ 100,\ 500)$ respectively, indicating convergence of the shift towards the asymptotic value $\frac{\alpha-1}{2}=0.34$. Other values of $M$ and $\nu$ show a similar behaviour.

Turning to the SW2 models, we use the form factor Eq.~(\ref{FF2}) to write the spectrum as
\be
\left(\frac{2-q}{\mu_n}+\frac{J_{q-1}(\mu_n)}{J_{q}(\mu_n)}\right)=
\tan \left(c_\beta\, \mu_n^\frac{1}{2\beta}
\right)\,,
\label{evsw2}
\ee
where $\mu_n=m_n/\rho= m_n z_1$.
Notice that for $\beta<\frac{1}{2}$, as we are assuming here, the left hand side of Eq.~(\ref{evsw2}) can be neglected for large masses and we simply obtain asymptotically
\be
m_n=\left(\frac{\pi\, n}{c_\beta}\right)^{2\beta}\! \rho \,.
\label{approxspectrumsw2}
\ee
Notice that $c_\beta^{-2\beta}$ is an $\mathcal O(1)$ constant. 
Let's compare this with the exact eigenvalues. We will restrict to the case $\beta=\frac{1}{4},\ M=0$ which allows us to determine the spectrum analytically.\footnote{The value $\beta=\frac{1}{4}$ corresponds to the interesting case of linear "Regge trajectories", i.e., $m_n^2\sim n$.} In fact there exists an exact solution to the massless scalar EOM in the background Eq.~(\ref{type2}) in terms of the confluent hypergeometric function $U(a,b,z)$,
\be
\psi_n^{\rm reg}(z)=U\left(-\frac{m_n^2}{12\rho^2},-1,3\rho z\right)\,,
\ee
where as before the superscript "reg" indicates that the solution regular at the singularity has been chosen. 
We can find the mass eigenvalues if $\psi_n$ satisfies a Dirichlet BC at the UV brane,
\be
U\left(-\frac{\mu_n^2}{12},-1,0\right)=-\frac{\Gamma(\frac{\mu_n^2}{12}-1)}{\pi} \sin\left(\pi \frac{\mu_n^2}{12}\right)=0\,,
\ee
and hence we obtain the exact spectrum
\be
m_n=\sqrt {12\, n}\,\rho\,,\qquad n\geq 2\,.
\label{exactsw2}
\ee
Since $c_{\frac{1}{4}}=\frac{\pi}{12}$ the spectra determined in Eqns.~(\ref{approxspectrumsw2}) and (\ref{exactsw2}) conincide precisely. The asymptotic spectrum thus turns out to be exact in this case (i.e.~there are no corrections for smaller $n$).

\section{Brane potentials from Soft Walls}
\label{sb}

Let us now imagine that we add a nontrivial bulk potential $V(\psi)$ for the scalar field. The holographic procedure outlined in Ref.~\cite{Witten:1998qj} shows that one can obtain the nonlinear boundary action at a hypersurface at $y=y_1$ perturbatively with the help of the bulk-to-boundary propagator $K(x,x',y)$ already calculated. Here we will follow a simplified approach that makes use of the assumption that the SW background is dominated by the radion $\phi$ and the backreaction of the additional scalar field $\psi$ is small. In particular, $\psi(y)\approx\psi(y_1)$ over the thickness of the soft wall, and the effective brane potential 
$\lambda_1(\psi)$
can be obtained from Eq.~(\ref{holo}) by a simple integration over the background:
\be
\sqrt{-g(y_1)}\lambda_1(\psi(y_1))=-V(\psi(y_1))\int_{y_1}^{y_s} dy\,\sqrt{-g(y)} \,,
\ee 
leading to
\be
\lambda_1(\psi)=-\frac{\nu^2}{\nu^2+4} \,V(\psi)\,\Delta y+\mathcal O(\Delta y^2)\,.
\label{IRpot}
\ee
Here, we have used the explicit forms of the SW metrics near the singularity, Eq.~(\ref{sw1pd}) and Eq.~(\ref{sw2pd}).
This surprisingly simple result can be obtained in a slightly more rigorous manner by looking at the nonlinear equation for the background:
\be
\psi''-4A'\psi'-V'(\psi)=0\,.
\ee
From this, we can write the regular solution near the singularity as
\be
\psi(y)=\psi_s+\frac{\nu^2}{2(\nu^2+4)}V'(\psi_s)(y_s-y)^2+\mathcal O((y_s-y)^3)
\ee
where $\psi_s=\psi(y_s)$ is the remaining integration constant. Imagine now that we know a solution in the bulk for $y<y_1$, maybe by making another approximation valid for small $y$. Matching the two solutions by imposing continuity of the solution and its derivative at $y_1$, one  obtains after elimination of $\psi_s$:
\be
\psi'(y_1)=-\frac{\nu^2}{\nu^2+4}V'(\psi(y_1))\,\Delta y+\mathcal O(\Delta y^2)\,.
\ee
The right hand side is equal to $\lambda_1'(\psi)$, so this is precisely the IR boundary condition we would have obtained by putting the brane potential, Eq.~(\ref{IRpot}), at $y=y_1$.

\section{Conclusions}
\label{concl}

In this note we have studied SW's by integrating over a small region near the singularity, typically present in such compactifications. We have classified SW's into two categories, SW1 and SW2, and argued that these already exhaust all backgrounds which fulfill two conditions
\begin{itemize}
\item
They lead to a mass gap. 
By the term mass gap we refer to the feature in the spectrum that the mass of the first heavy mode is separated by a gap from zero. We do not exclude in the definition the possibility that there is, in addition, an isolated exact zero mode, nor that the spectrum above the first massive mode is continuous. 
\item
They do not create spurious contributions to the cosmological constant at the singularity: the ansatz for the metric is 4d flat and hence the 4d cosmological constant must vanish by an appropriate tuning between bulk and UV brane, $V_{4d}(\phi_0)=0$. However, there is a hidden contribution to the cosmological constant from the singularity \cite{Forste:2000ps, Cabrer:2009we} which leads to the condition Eq.~(\ref{condcc}).
\end{itemize}
Imposing these constraints leads to the singularities given in Eqns.~(\ref{sw1pd}) to (\ref{sw2cf}).
We have shown that such SW's can be given a physically equivalent description in terms of IR branes. The IR brane Lagrangian can be obtained by holographic projection onto a hypersurface located at $y_1$ smaller but close to $y_s$, the location of the singularity. 

There are several advantages of this method:
\begin{itemize}
\item
The universal nature of the SW's is reflected in the IR brane Lagrangian for $y_1$ sufficiently close to $y_s$.
\item
It cleanly separates the bulk physics from the physics of the singularity, thus facilitating a comparison of SW's to the standard two-brane compactifications.
\item
It gives a meaningful and trustable IR brane Lagrangian even in the energy range above the IR scale. 
\item
It defines a useful approximation scheme: The "new" bulk, defined by $0<y<y_1$, can often be approximated by a pure RS metric.
\end{itemize}
Technically, the last point represents a huge simplification as exact SW backgrounds are notoriously hard to treat analytically.
As an application and an illustration of the above points, we have calculated the spectrum for a scalar field in the two classes of SW's. The numerical accuracy is very good. As expected, it improves for larger KK masses, since the subleading terms of the SW metric near $y_s$ become less important.

There are several extensions and applications of this work. Firstly, the generalization to fields of nonzero spin, which is straightforward. For gauge bosons, the form factors can be computed in complete analogy
to the scalar case, by simply making the substitution $4A'\to 2A'$ and setting the mass term to zero. Fermionic form factors are more subtle due to the fermions obeying coupled first order equations, some details can be found in App.~A.
For gauge fields and gravity, gauge and diffeomorphism invariances allow one to deduce even terms of higher than quadratic order from the purely bilinear terms, opening up the possibility to study scattering of, say, graviton KK modes at energies above the IR scale. 

\section{Acknowledgements}

It is a pleasure to thank J.~A.~Cabrer and M. Quir\'os for many useful discussions.
This work was supported in part by the European Commission under contracts
PITN-GA-2009-237920, the ERC Advanced Grant 226371
(``MassTeV''), by ANR (CNRS-USAR) contract 05-BLAN-007901 and by CNRS PICS no. 3747
and 4172.

\appendix
\section{Fermions}

In this appendix we briefly comment on the fermionic case. Let us consider the action
\begin{align}
S=& \int dy\,e^{-3A}\left(i\bar\psi_L\, /\hspace{-.22cm}\partial\,\psi_L
+i\bar\psi_R\, /\hspace{-.22cm}\partial\,\psi_R\right)\nonumber\\
&+
e^{-4A}\,\left(
\bar\psi_R\psi_L'+2A'\,\bar\psi_R\psi_L
-M(y)\bar\psi_R\psi_L+{\rm h.c.}\right)\,.
\label{A1}
\end{align}
where for sake of generality we have allowed the bulk mass to depend on $y$.
Defining the kernels by 
\bea
\psi_{L}(p,y)=e^{2A(y)-2A(y_1)} \frac{K_{L}(p,y)}{K_L(p,y_1)} \psi_{L}(p,y_1)\,,\\
\psi_{R}(p,y)=e^{2A(y)-2A(y_1)} \frac{K_{R}(p,y)}{K_R(p,y_1)} \psi_{R}(p,y_1)\,,
\eea
we can rewrite the Dirac equation as \cite{Contino:2004vy}
\bea
\sqrt{-p^2}K_{L,R}&=&e^{-A}(M\pm \partial_y)K_{R,L}\,,\label{Dirac2}\\
\,/\hspace{-.23cm}p\,\psi_R(p,y_1)K_L(p,y_1)&=&\sqrt{-p^2}\psi_L(p,y_1)K_R(p,y_1)\,.
\eea
From Eq.~(\ref{holo}), the SW action can be written as
\bea
S_{SW}&=&-\int\frac{d^4p}{(2\pi)^4} e^{-4A(y_1)}\bar
\psi_R(p,y_1)\psi_L(p,y_1)\nonumber\\
&=&
\int\frac{d^4p}{(2\pi)^4} e^{-4A(y_1)}\bar\psi_L(p,y_1)
\mathcal F(p)
\psi_L(p,y_1)
\eea
where the form factor is
\be
\mathcal F(p)=-\frac{\,/\hspace{-.23cm}p\,K_R(p,y_1)}{\sqrt{-p^2}K_L(p,y_1)}
\ee
As a matter of fact, rewriting Eq.~(\ref{Dirac2}) as a second order equation in conformally flat coordiantes, we obtain the Schr\"odinger-like equation
\be
-p^2 K_{L,R}(z)=
(\tilde M^2\pm  \tilde M')K_{L,R}-K_{L,R}''\,,\qquad \tilde M=e^{-A}M
\ee
For a constant mass $M$, this equation has a gapless continuum as a spectrum (as $\tilde M$ goes to zero at the singularity), and hence does not provide us with a very interesting example. Let us instead consider the example where $M=-\frac{1}{2}W$, where $W$ is the superpotential generating the SW. Then $K_L$ follows the equation
\be
-p^2 K_{L}(z)=
\left(\frac{1}{4}A'(z)^2-\frac{1}{2}A''(z)\right)K_{L}-K_{L}''
\ee
This is the equation of motion for a gauge boson in the corresponding SW background, and one can speculate that the particular choice for $M$ follows from a supersymmetrization of the theory. The calculation for the form factors is very similar to the case of the scalars and for this reason we will not present it here.

\end{document}